\documentclass[11pt]{article}

\usepackage{hyperref}
\hypersetup{
    colorlinks=true,
    linkcolor=blue,
    filecolor=magenta,    
    urlcolor=blue,
}

\usepackage{color}
\usepackage{latexsym}
\usepackage{epsfig,graphics}
\usepackage{amsmath,amssymb,amsthm}
\usepackage{graphics,epsfig,calc}
\usepackage{bm}

\usepackage{upref}
\usepackage{setspace}
 \newcommand{\beqn}{\begin{eqnarray}}
 \newcommand{\eeqn}{\end{eqnarray}}
 \newcommand{\be}{\begin{equation}}
 \newcommand{\ee}{\end{equation}}
 \newcommand{\ba}{\begin{array}}
 \newcommand{\ea}{\end{array}}

 \newcommand{\pa}{\partial}
 \newcommand{\re}{\ref}
  \newcommand{\ci}{\cite}
 \newcommand{\ds}{\displaystyle}
 \newcommand{\la}{\label}

   \newcommand{\blangle}{{\big\langle}}
  \newcommand{\brangle}{{\big\rangle}}

 \newcommand{\fr}{\frac}

\newcommand{\ov}{\overline}

\newcommand{\cF}{{\cal F}}

\newcommand{\bA}{{\bf A}}
\newcommand{\bba}{{\bf a}}

\newcommand{\cE}{{\cal E}}
\newcommand{\bF}{{\bf F}}

\newcommand{\bH}{{\bf H}}
\newcommand{\cI}{{\cal I}}

\newcommand{\p}{{\bf p}}
\newcommand{\cP}{{\cal P}}
\newcommand{\bS}{{\bf S}}
\newcommand{\cW}{{\mathbb{W}}}
\newcommand{\cY}{{\mathbb{Y}}}

\newcommand{\ve}{\varepsilon}
\newcommand{\vp}{\varphi}
\newcommand{\De}{\Delta}
\newcommand{\de}{\delta}
\newcommand{\dv}{{\rm div}}

\newcommand{\al}{\alpha}
\newcommand{\om}{\omega}
\newcommand{\na}{\nabla}
\newcommand{\bPi}{{\bm\Pi}}
\newcommand{\bpi}{\bm\pi}
\newcommand{\Si}{\rm \Sigma}

\newcommand{\R}{\mathbb{R}}

\newtheorem{theorem}{Theorem}[section]

\newtheorem{defin}[theorem]{Definition}

\newtheorem{lemma}[theorem]{Lemma}
\newtheorem{remark}[theorem]{Remark}

\newtheorem{cor}[theorem]{Corollary}
\newtheorem{pro}[theorem]{Proposition}
\newcommand{\bp}{\begin{pro}}
\newcommand{\ep}{\end{pro}}

\newcommand{\bt}{\begin{theorem}}
\newcommand{\et}{\end{theorem}}

\newcommand{\bl}{\begin{lemma}}
\newcommand{\el}{\end{lemma}}

\newcommand{\bc}{\begin{cor}}
\newcommand{\ec}{\end{cor}}

\newcommand{\const}{\mathop{\rm const}\nolimits}

\newcommand{\bpr}{\begin{proof}}
\newcommand{\epr}{\end{proof}}

\newcommand{\br}{\begin{remark}}
\newcommand{\er}{\end{remark}}

\newcommand{\bd}{\begin{defin}}
\newcommand{\ed}{\end{defin}}

\begin{document}
\begin{center}
{\Large\bf On orbital stability of  solitons for 2D Maxwell--Lorentz  equations}
 \bigskip \medskip

 {\large A. I. Komech
 \footnote{ 
 Supported partly by Austrian Science Fund (FWF) PAT 3476224}
 \\ 
\centerline{\it Institute of Mathematics,  BOKU University,
Vienna}


\centerline{alexander.komech@gmail.com}
}

\medskip

 {\large E.A. Kopylova}
 \footnote{ 
 Supported partly by Austrian Science Fund (FWF) P34177
 }
 \\
{\it Faculty of Mathematics,   Vienna University
  }

\end{center}

\begin{abstract}

We prove the orbital stability of soliton solutions  for 
2D Maxwell--Lorentz system with extended charged  particle.
The solitons  corresponds to the uniform motion and rotation of the particle.

We reduce the  corresponding Hamilton system  by the canonical transformation via transition to a comoving frame. 
The solitons are the critical points of the reduced Hamiltonian. 
The key point  of the proof is a lower bound  for the  Hamiltonian.
\end{abstract}

\newpage

\section{Introduction}
We consider the 2D Maxwell--Lorentz equations with extended charged 
spinning particle
\be\la{mls2}
\left\{\ba{rcl}
\dot E(x,t)\!&\!=\!&\!J\na B(x,t)-[\dot q(t)-\om(t) J(x-q(t))] \rho(x-q)
\\
\dot B(x,t)&=&-\na \cdot JE(x,t),\quad
\na\cdot E(x,t)=\rho (x-q(t)),
\\
m\ddot q(t)\!&\!=\!&\!\blangle E(x,t)+ [J\dot q(t) +\om(t) (x-q(t)) ]B(x,t),\rho(x-q(t))\brangle
\\
 I\dot \om(t)\!&\!=\!&\!\blangle (x-q(t))\cdot [J E(x,t)-\dot q(t) B(x,t) ],\rho(x-q(t))\brangle
\ea\right|,~~x\in\R^2, ~~t\in\R.
\ee
Here $J=\begin{pmatrix}
0&1
\\
-1&0
\end{pmatrix}$,  $E(x,t)\in\R^2$ is the electric field, $B(x,t)\in\R$ is the magnetic field, $\rho(x-q(t))$ is the charge distribution of the extended
particle centered at a point $q(t)\in\R^2$,  $m$ is the mass of the particle,  {\color{red}$I>0$} is  its moment of inertia,
 and $\om(t)=\dot\vp(t)$ is the angular velocity of the particle rotation. The brackets $\langle,\rangle$ denote the inner product in the Hilbert space $L^2:=L^2(\R^2)\otimes\R^2$.  We choose the units where the speed of light is $c=1$. 

Note that the 2D Maxwell--Lorentz equations  can be introduced directly from 3D Maxwell--Lorentz system 
\ci[(2.39)--(2.41)]{S2004} by substituting $x=(x_1,x_2,0)$ and 
$$
E(x,t)=(E_1(x,t),E_2(x,t),0),\quad B(x,t)=(0,0,B_3(x,t)),\quad q(t)=(q_1(t),q_2(t),0).
$$
In the main body of the paper, we focus on the equations with a {\it nonrelativistic  particle}, 
and at the end, we comment on the corresponding modifications for the case of a {\it relativistic particle}. 

We assume that the charge density $\rho(x)$ is smooth and  rotation-invariant, i.e.,
\be\la{rosym}
\rho\in C_0^\infty(\R^2),\qquad\rho(x)=\rho_1(|x|).
\ee
Moreover, we assume that the total charge equals zero (neutrality of the particle):
\be\la{rho02}
\int\rho(x)dx=0.
\ee
The Maxwell potentials $A(x,t)=(A_1(x,t),A_2(x,t))$, $\Phi(x,t)$ are introduced by
\be\la{AA}
B(x,t)=\na\cdot JA,\qquad E(x,t)=-\dot A(x,t)-\na \Phi(x,t).
\ee
We choose the Coulomb gauge
$\na\cdot A(x,t)=0$. The second line of (\ref{mls2}) implies that $\Delta\Phi(x,t)=-\rho(x-q(t))$. Hence,
\be\la{2mA3}
\Phi(x,t)=\Phi_0(x-q(t)),~~{\rm where} \qquad \Phi_0(y)=-\fr1{2\pi}\int \log |z-y| \rho(z)dz.
\ee
In the Maxwell potentials, the system  (\ref{mls2}) becomes
\be\la{2ml2}
\left\{\ba{rcl}
\ddot A(x,t)&=&\De A(x,t)+\cP\big([\dot q(t)-\dot\vp(t) J(x-q(t))]\rho(x-q)\big)
\\
m\ddot q(t)&=&\blangle-\dot A(x,t)+ [J\dot q(t) +\dot\vp(t) (x-q(t))]\na\cdot JA(x,t),\rho(x\!-\!q(t)) \brangle
\\
 I\ddot \vp(t)\!&=&\!\!\blangle (x\!-\!q(t))\cdot [-J \dot A(x,t)\!-\dot q(t) \na\cdot JA (x,t) ],\rho(x\!-\!q(t))\brangle
\ea\right|.
\ee
Here  ${\cal P}$ is the projection onto the space of solenoidal (divergence-free) vector fields.
In the Fourier space,
\be\la{Pi-e}
\widehat{{\cal P}a}(k)=\hat a(k)-\frac{\hat a(k)\cdot k}{k^2}k.
\ee
The system is invariant under the action of
the group of translations and rotations of the space $\R^2$.
Accordingly, the system  admits solitons of the form
\be\la{solvom}
S_{v,\om}(t)=
(A_{v,\om}(x-vt),\dot A_{v,\om}(x-vt), vt, v,\om )
\ee
moving with constant speed $v$ and rotating with constant angular velocity $\om$.

The main result of present paper is the orbital stability of the solitons. 
We establish the Lagrangian and the Hamiltonian structures of the system (\ref{mls2}), and prove the conservations 
of  energy, linear and angular momenta.

The proof of the orbital stability of solitons relies on  the reduction 
of the  system (\ref{2ml2})  by the canonical transformation via transition to  a comoving frame.
The Hamiltonian of the reduced system depends on two parameters:
the total linear momentum $P$ and the 
angular momentum $M$, which are invariant in time. 

The key point is the proof of the lower bound  for the reduced Hamiltonian
with the fixed $P$ and $M$.
The bound implies that the soliton is a strict global minimizer of the reduced Hamiltonian.
 We show that this fact implies the orbital stability. 
Moreover, the lower bound implies that  the reduced system admits a unique 
soliton for every $P,M\in\R^2$.
 
Let us comment on related results.

Low-dimensional models are widely exploited
as useful models  in the classical and quantum field theories. For example, 
't Hooft \ci{H1974} established the confining for quarks in the meson theory in 1+1 space-time. 
The 2D version of the Maxwell electrodynamics 
was considered by Schwinger in \ci{S1962}, 
 to show the possibility of the nonzero mass particles in the gauge theories.
In \ci{HK2000}, the Maxwell--Lorentz equations  in 1D space were considered
 in the context of plasma dynamics, when the charge and currents
correspond to a smooth distribution of the charged dust subject to the Lorentz dynamics. 
The well-posedness is proved and the point charge asymptotics is analyzed.
The solitons of 2D Maxwell--Lorentz equations and their stability were not consider previously.

 In \ci{BG1993}, Bambusi and Galgani proved the existence and  orbital stability of solitons with velocities $|v|<1$ for 
 the 3D Maxwell--Lorentz system  without spinning. The proof relies on the transition to a comoving frame 
and a lower bound for the reduced Hamiltonian.
The global convergence to solitons for the same system was proved in \ci{IKM2004}.
In \ci{KS2000}, the global attraction to stationary states was established  for a similar system  with a
 relativistic particle in presence of an external confining potential. 

 The results \ci{IKM2004, K2010,KS2000}  provide the first rigorous proof of the radiation damping
 in classical electrodynamics.  All these results were obtained  under the {\it Wiener-type condition} on the charge density $\rho$.
 For the corresponding  surveys, see \ci{KK2020} and \ci{KK2022}.

The global convergence to rotating solitons was established  in \ci{IKS2004} for solutions of the 3D Maxwell--Lorentz system
 with $q(t)\equiv 0$ in the case of sufficiently small charge density $\rho$.
This result was strengthened in \ci{K2010} under a considerably weaker  Wiener-type condition.

 In \ci{IKS2011},   the asymptotic completeness  was proved for the scattering of solutions to the 
 Maxwell--Lorentz equations without spinning. The result refers to solutions which are close to a solitary manifold.
 
  In \ci{KK2023,KK2024}  the orbital  stability of  solitons was studied for the 3D Maxwell--Lorentz system with moving and spinning particle.
  The solitons move with a velocity $v\in\R^3$, $|v|<1$ and rotate with 
  an angular velocity $\om\in\R^3$.
  The results are obtained for solitons either with $\om=0$ or with $v=0$.
  The proofs rely on the  Hamiltonian--Poisson structure of the equations 
  which has been constructed in \ci{KK2023_LH}.
  
\setcounter{equation}{0}
\section{Well-posedness}\la{WA}
Let us introduce a phase space for the system \eqref{2ml2}. Denote by 
$\dot H^1$  the closure of $C_0^\infty (\R^2)\otimes \R^2$ with the respect to the norm 
$\Vert A\Vert_{\dot H^1}=\Vert \nabla A\Vert_{L^2}$, where $L^2:=L^2(\R^3)\otimes\R^2$.
Let $\cF^0$, $\dot\cF^1$ be the subspaces of solenoidal vector fields:
\be\la{FF} 
\cF^0:=\{A\in L^2: \dv A(x)\equiv 0\},\qquad 
\dot\cF^1:=\{A\in \dot H^1: \dv A(x)\equiv 0\}.
\ee
Denote
\be\la{cWH}
\cF:=\dot\cF^1\oplus \cF^0,\quad \cY:=\dot\cF^1\oplus\cF^0\oplus \R^2\oplus \R^2\oplus \R^2,\quad
\cW:=L^2\oplus H^{-1}\oplus\R^2\oplus \R^2\oplus \R^2.
\ee
\bp\la{pwpA}
{\rm i)} For any initial state 
$Y(0)=(A(x,0),\dot A(x,0),q(0), \dot q(0),\vp(0),\dot\vp(0))\in\cY$, there exists a unique
solution of the system {\rm (\ref{2ml2})}, 
\be\la{soluY}
Y(t)=(A(x,t),\dot A(x,t),q(t),\dot q(t),\vp(t),\dot\vp(t))\in C(\R,\cY)\cap C^1(\R,\cW).
\ee
{\rm ii)} The map 
\be\la{cont}
W(t): Y(0)\mapsto Y(t)
\ee
 is continuous in $\cY$ for every $t\in\R$.
\smallskip\\
{\rm iii)} For $Y(0)\in\cY$, the energy $\cE(t)$,  linear momentum $P(t)= (P_1(t), P_2(t))$ and angular momentum $M(t)$  are conserved:
\beqn\la{ecoA}
\cE(t):&=&\fr12\int[\dot A^2(x,t)+|\na A(x,t)|^2]dx+\fr 12m\dot q^2(t)
+\fr12I\dot \vp^2(t)=\const,\\
\la{mcoA}
P_j(t)&:=&-\blangle \dot A(x,t), \na_jA(x,t)\brangle+ m\dot q_j(t)+\blangle\rho(x-q(t)), A_j(x,t)\brangle=\const,~j=1,2,\\
\la{amcoA}
M(t)&:=&I\dot\vp(t)-\blangle [J(x-q(t))]\cdot A(x,t),\rho(x-q(t))\brangle=\const.
\eeqn
{\rm iv)} Let 
\be\la{AP0}
\left\{\ba{l}
A(x,0)\in C^3(\R^2)\!\otimes\!\R^2,\qquad \dot A(x,0)\in C^2(\R^2)\!\otimes\!\R^2
\\
A(x,0)=0,\,\,\dot A(x,0)+\na\Phi_0(x-q(0))=0,\quad |x|>R_0
\ea\right|
\ee
for some $R_0>0$. Then
\be\la{AP1}
\left\{\ba{l}
A(x,t)\in C^2(\R^2\times\R)\!\otimes\!\R^2,
\\
  |\pa_x^\al A(x,t)|+|\pa_x^\al \dot A(x,t)|\le C_\al(1+|x|)^{-2-|\al|},
  \quad |x|>R(t)+1,\qquad \forall\al
\ea\right|,
\ee
where 
$R(t)=R+|q(0)|+\ov vt+R_\rho$ with 
$\ov v>0$ defined by the relation $\cE=\fr12 m\ov v^2$.
\ep
\bpr
The proof  is similar to that of  Proposition A.5 in \ci{KS2000}.
 Namely, consider at first the case of  smooth initial  functions  $A(x,0), \dot A(x,0)$ with compact supports.
 The elimination of the fields reduces
 the system (\ref{2ml2}) to  the system of nonlinear integral equations  for $q(t)$ and $\om(t)$. 
 Then the existence and uniqueness of the solution 
 $(q(t),\om(t))$
 for small $|t|$  follows by the
  application of the contraction mapping principle.
  The corresponding fields $A(\cdot,t), \dot A(\cdot,t)$ are smooth and have compact supports, hence, the energy conservation
  (\ref{ecoA})
   follows by standard
  integration by parts. Therefore,  the solution $Y(t)$ can be extended to all $t\in\R$. 
 
  The continuity of the  map 
  $W(t)$, as constructed
  on the dense subspace of
   $\cY$, follows from the continuity 
  of the map $\cY\to C^2([0,t]; \R^2\oplus\R^2)$ defined
  as $Y(0)\mapsto (q(\cdot),\om(\cdot))|_{[0,t]}$.
  Now 
  i)  and (\ref{ecoA}) follow
  for general initial state $Y(0)\in \cY$
  by suitable approximations of $Y(0)$.
   The conservation laws (\ref{mcoA}), and (\ref{amcoA}) will be proved later. 
 
 It remains to prove iv).
 Note that  $q(\cdot)\in C^3(\R)\otimes\R^2$ 
 and $\om(\cdot)\in C^2(\R)\otimes\R^2$
 by (\ref{soluY})
 and the last two equations of (\ref{2ml2}). This fact and 
 the first line of 
 (\ref{AP0}) imply
 the first line of (\ref{AP1})
  due to the Kirchhoff integral representation  of solutions
   of the first equation of (\ref{2ml2}).
 Further, 
 applying $-J\nabla$ to both sides of the first equation in  (\ref{AA}), and using $\na\cdot A=0$,
 we obtain that 
 $$
 \De A(x,t)=-J\nabla B(x,t).
 $$
 The second line of (\ref{AP0}) implies that
 \be\la{EB0}
 E(x,0)=B(x,0)=0,\qquad |x|>R. 
 \ee
  Moreover,
  (\ref{ecoA}) implies that
  $|\dot q(t)|\le \ov v$, $t\in\R$.
  Hence, $|q(t)|\le |q(0)|+\ov v|t|$.
   
 Therefore, (\ref{EB0}) implies that
 $E(x,t)=B(x,t)=0$, for $|x|>R(t)$ 
  by the 2D analogue of the integral representations (A.4) from \cite{KS2000}.
  Hence,
\be\la{AAA}
A(x,t)=-\fr1{2\pi}\int_{|y|<R(t)} J\nabla B(y,t)\log|x-y|dy 
=-\fr1{2\pi}J\nabla \int_{|y|<R(t)} B(y,t)\log|x-y|dy.
\ee
Now the second line of (\ref{AP1}) follows.
\epr
\setcounter{equation}{0}
\section{The Lagrangian structure}
Here we show that,  under  assumption (\ref{rosym}),  sufficiently smooth solutions
$X(t)=(A(t),q(t),\om(t))$ of the 2D system \eqref{2ml2} satisfy the Hamilton least action  principle.
The Lagrangian $L$ is well known  (cf. \ci[(28.6)]{LL1975}:
 \be\la{LagRA}
L(A,\dot A,q,\dot q,\vp,\dot\vp)
=
\fr12\int[\dot A^2-(\na A)^2]dx+\fr {m\dot q^2}2+\fr{I\dot\vp^2}2
+
\blangle [\dot q-\dot\vp J(x-q)]\cdot A,\rho(x-q)\brangle
\ee
up to an additive constant depending on $\Phi$.
This Lagrangian  is well defined 
and  Fr\'echet differentiable 
on the extended Hilbert  phase space $\cY$.
\bp \la{LS}
Under the assumption (\ref{rosym}), the system (\ref{2ml2})
with $\om(t)=\dot\vp(t)$   is equivalent to the Euler--Lagrange equations
\be\la{EL}
\dot \Pi =D_A  L, \qquad 
\dot p =D_q  L, \qquad \dot M =D_\vp  L,
\ee
where the canonical conjugate momenta are defined by
\be\la{cam}
\Pi:=D_{\dot A} L,  \qquad 
p:=D_{\dot q} L, \qquad M:=D_{\dot\vp} L.
\ee
\ep
\begin{proof}
All the calculations of  variations are justified by 
$C^2$-approximations, which exist by Proposition \ref{pwpA}, iv).

{\it Step i)} Differentiating (\ref{LagRA}), and using (\ref{AA}), we find
the momenta
\be\la{pA}
\Pi=
\dot A,\qquad
p=m\dot q+\blangle A(q+y),\rho(y)\brangle,
\qquad M=I\dot\vp-\blangle [Jy]\cdot A(q+y),\rho(y)\brangle.
\ee
It remains to calculate the  derivative $D_A L$
on the right-hand side 
of the first equation (\ref{EL}). 
For any  $h\in \dot\cF^1$, we have
$$
\fr12\blangle D_A \int (\na A(x))^2dx, h\brangle=
\fr12 \fr d{d\ve}\Big|_{\ve=0} \int |\na[A(x)+\ve h(x)]|^2dx
=  \!\int \na A(x)\na h(x) dx
=-\blangle\De A,h \brangle.
$$
Hence, 
\be\la{pA3}
D_A  L=\De A(q+y)+\cP \big([\dot q-\dot\vp Jy]\rho(y)\big).
\ee
Now the first equation of (\ref{EL}) coincides with the Maxwell equation (the first line  of (\ref{2ml2})).

{\it Step ii)}  Now we check the Newton equation (the  second equation of (\ref{2ml2})).
Differentiating  (\ref{LagRA}) in $q_k$, we obtain
\beqn\la{pq}
D_{q_k}  L&=&D_{q_k}
\blangle [\dot q-\dot\vp Jy]\cdot A(q+y),\rho(y)\brangle
=\blangle[\dot q-\dot\vp Jy]\cdot \na_k A(q+y),\rho(y)\brangle,~~ k\!=\!1,2.
\eeqn
Hence, the corresponding Euler--Lagrange equation reads as the system
\be\la{eq2}
m\ddot q_k+\blangle\dot A_k(q+\!y,t),\rho(y)\brangle
+\blangle(\dot q(t)\!\cdot\!\na) A_k(q+\!y,t),
\rho(y)\brangle
\!=\!\blangle[\dot q\!-\!\dot\vp Jy]\cdot \na_k A(q+\!y,t),\rho(y)\brangle.  
\ee
We should check that this system coincides with the second equation of  (\ref{2ml2})
with $\om(t)=\dot\vp(t)$:
\be\la{mdo}
m\ddot q_k(t)=\blangle-\dot A_k(q+y,t)+[(J\dot q)_k(t) +\dot\vp(t) y_k]\na\cdot JA(q+y,t)
,\rho(y) \brangle.
\ee
It remains to prove that
\be\la{mdo2}
-\blangle(\dot q\cdot\na) A_k,\rho\brangle
+\blangle[\dot q-\dot\vp Jy]\cdot \na_k A,\rho\brangle
=\blangle[(J\dot q)_k +\dot\vp y_k]\na\cdot JA ,\rho \brangle,~~ k\!=\!1,2.
\ee
It suffices to consider the case $k=1$. Now (\ref{mdo2}) becomes
$$
-\blangle(\dot q_1\na_1+\dot q_2\na_2) A_1,
\rho\brangle
+\blangle[\dot q_1-\!\dot\vp y_2]\na_1 A_1,\rho\brangle+[\dot q_2+\!\dot\vp y_1]\na_1 A_2,\rho\brangle
=\blangle[\dot q_2 +\dot\vp y_1]\na\cdot JA,\rho \brangle.
$$
After cancellations, this identity is equivalent to
\beqn\la{mdo4}
&&
\dot\vp\blangle [- y_2\na_1+y_1\na_2]A_1,\rho\brangle
=\dot\vp\blangle A_1,[ y_2\na_1-y_1\na_2]\rho\brangle=0.
\eeqn
This identity follows  by the rotational symmetry (\ref{rosym}).
 
{\it Step iii)} Finally, let us check that the last equation of (\ref{EL}) is equivalent to the last  (torque) equation of the system (\ref{2ml2}). 
The right-hand side of the last equation of (\ref{EL}) 
vanishes since the Lagrangian (\ref{LagRA}) does not depend on $\vp$.
Now,  the expression for $M$ from (\ref{pA}) implies
\be\la{cam3}
\dot M(t)=
I\ddot\vp(t)-\blangle[Jy]\cdot [\dot A(q+\!y,t)
+(\dot q\cdot\na) A(q+\!y,t)],\rho(y)\brangle=0.
\ee
So,  the angular momentum is conserved:
\be\la{L}
M(t)=M,\qquad t\in\R.
\ee
To prove the equivalence
 of the last equations of (\ref{EL}) and  (\ref{2ml2}),
we should check that
\be\la{cam32}
-\blangle (y_1\dot q_1+y_2\dot q_2)(\na_1A_2-\na_2A_1),\rho\brangle=
\blangle y_2[ (\dot q_1\na_1+\dot q_2\na_2)A_1]-y_1[(\dot q_1\na_1+\dot q_2\na_2)A_2],\rho\brangle.
\ee
After cancellations, we arrive at
$$
\dot q_2
\blangle (-y_2\na_1+y_1\na_2)A_2,\rho\brangle=
\dot q_1\blangle (-y_1\na_2+y_2\na_1)A_1,\rho\brangle.
$$
Integrating by parts, and using the rotational symmetry (\ref{rosym}), we
conclude that both parts of the last equation vanish.
\end{proof}
\setcounter{equation}{0}
\section{The conservation of linear momentum}
The Lagrangian (\ref{LagRA}) is invariant under translations $x\mapsto x+a$
and rotations $R(\theta): x_1+ix_2\mapsto  e^{i\theta}(x_1+ix_2)$ of the plane $\R^2$.
The rotation symmetry implies the conservation 
(\ref{L}) of the angular momentum $M$.
Let us calculate the conserved linear momentum $P$.
Consider the action of  the group of translations 
$$
T_a:X=(A,q,\vp)\mapsto T_aX=
(A(x-a),q+a,\vp),\qquad a\!\in\!\R^2.
$$
The derivative of this map is
$$
T'_a:\dot X=(\dot A,\dot q,\dot\vp)\mapsto T_a'\dot X=
(\dot A(x-a),\dot q,\dot\vp).
$$
It is easy to check that $L(T_a X, T_a'\dot X)=L(X,\dot X)$.
Hence, we can apply
the Noether theorem on invariants \ci[p. 88]{A1989} to the one-parametric subgroups
$T_{\ve a}$ with  any $a\in\R^2$ and $\ve\in\R$.
It suffices to consider
 $C^2$-approximations, which exist by
Proposition \ref{pwpA}.
Thus,
we obtain  
the conservation  
of the projection of the linear momentum 
\beqn\la{mocop}
P(Y,\dot Y)\cdot a&=&D_{\dot Y}L\cdot\fr d{d\ve}\Big|_{\ve=0}T_{\ve a}Y=
(\dot A, m\dot q+\blangle\rho(x-q) ,A(x)\brangle, D_{\dot \vp}L)\cdot(-a\cdot\na A,a,0)
\nonumber\\
&=&
-\blangle \dot A, a\cdot\na A\brangle+ m\dot q\cdot a+\blangle\rho(x-q), A(x)\brangle\cdot a.
\eeqn
Therefore, the conserved linear momentum is the vector 
\be\la{moco}
P=-\blangle \Pi, \na_* A\brangle+ p=-\blangle \dot A, \na_* A)\brangle+ m\dot q+\blangle\rho(x-q), A(x)\brangle,
\ee
where the expression with $\na_*$ is defined by
\be\la{nA}
\blangle \Pi, \na_* A\brangle_j=\blangle\Pi,\na_j A\brangle,\quad j=1,2.
\ee
\setcounter{equation}{0}
\section{The Hamiltonian structure}
The Hamiltonian structure is obtained through the Legendre transformation
\ci{A1989,MR2002}
\be\la{Leg}
(\dot A,\dot q,\dot\vp)\mapsto(\Pi,p,M),\quad
H(A,q,\vp,\Pi,p,M)=\blangle \Pi,\dot A\brangle+p\dot q+M\dot\vp- L.
\ee
The formulas (\ref{pA}) give
\be\la{Apv}
\dot A=\Pi,\qquad \dot q=\fr1m (p-\blangle A(q+y),\rho(y)\brangle),\qquad
\dot\vp=\fr1I(M+\blangle [Jy]\cdot A(q+y),\rho(y)\brangle).
\ee
Hence,  we obtain the Hamiltonian
\beqn\la{Hp}
H\!\!&\!\!=\!\!&\!\!\fr12\int[\Pi^2+|\na A|^2]dy
+[m\dot q+\blangle A(q+y),\rho(y)\brangle]\dot q+
[I\dot\vp-\blangle [Jy]\cdot A(q+y),\rho(y)\brangle]
\dot\vp
\nonumber\\
\!\!&\!\!\!\!&\!\!-\fr {m\dot q^2}2-\fr{I\dot \vp^2}2
-\blangle [\dot q-\dot\vp Jy]\cdot A(q+y),\rho(y)\brangle
=\fr12\int[\Pi^2+|\na A|^2]dy+\fr {m\dot q^2}2+\fr{I\dot \vp^2}2,\qquad
\eeqn
which coincides with (\ref{ecoA}). 
Here $\ds\fr{m\dot q^2}2$ and $\ds\fr{I\dot \vp^2}2$ are {\it defined} according to (\ref{Apv}):
\be\la{dqp}
\ds\fr{m\dot q^2}2=\ds\fr1{2m} (p-\blangle A(q+y),\rho(y)\brangle)^2,
\qquad
\ds\fr{I\dot \vp^2}2=\ds\fr1{2I}(M+\blangle [Jy]\cdot A(q+y),\rho(y)\brangle)^2.
\ee
The Hamiltonian (\ref{Hp}) is well defined and Fr\'echet differentiable on the Hilbert
phase space $\cY$.
The system (\ref{2ml2}) is equivalent to the canonical Hamiltonian system
\be\la{canH}
\dot A=D_\Pi H,\,\,\,\dot \Pi=-D_A H;\quad\dot q=D_{p} H,\,\,\,\dot p=-D_q H;
\quad\dot \vp=D_{M} H,\,\,\,\dot M=-D_\vp H.
\ee
Now the energy conservation (\ref{ecoA}) obviously follows.
\setcounter{equation}{0}
\section{The canonical transformation to a comoving frame}

Let us denote the fields in the ``comoving frame'' as
\be\la{canA}
\bA(y)=A(q+y), \qquad \bPi(y)=\Pi(q+y),
\ee
and define the map of the phase space $\cY\oplus\R$,
\be\la{cat}
T:(A,\Pi,q,p,\vp,M)\mapsto (\bA,\bPi,q,P,\vp,M),
\ee
where $P$ is defined by (\ref{moco}).
Let us show that $T$ is a canonical transformation, i.e. it preserves the canonical form
(\ci{A1989,MR2002}):
\be\la{canf}
\blangle \bPi,\dot \bA \brangle+P\dot q+M\dot\vp=
\blangle \Pi,\dot A\brangle+p\dot q+M\dot\vp.
\ee
Indeed, 
 differentiating in time $\bA(y,t):=A(q(t)+y,t)$, we obtain
\be\la{Tca2}
\dot \bA(y,t)=
 \dot A(q+y,t) +\dot q(t)\cdot\na  A(q+y,t).
\ee
 Hence, the identity (\ref{canf}) reduces to
$\blangle  \Pi,\dot q\cdot\na  A\brangle+P\cdot\dot q=p\dot q,$
which holds due to (\ref{moco}).

As a result, the  canonical Hamiltonian equations (\ref{canH})
in the variables $(A,\Pi,q,P,\vp,M)$ with the Hamiltonian (\ref{Hp})
are equivalent to the similar equations in the variables 
$(\bA,\bPi,q,P,\vp,M)$ with the Hamiltonian
\beqn\la{HT}
\bH(\bA,\bPi,P,M)&=&\fr12\int[|\bPi|^2+|\na \bA|^2]dy+
\fr1{2m} (P+\blangle \bPi, \na_*  \bA\brangle-\blangle \bA,\rho\brangle)^2
\nonumber\\
&&
+\fr1{2I}(M+\blangle [Jy]\cdot \bA(y),\rho(y)\brangle)^2,
\eeqn
where we used the expressions  (\ref{dqp}), (\ref{moco}) 
 and (\ref{L}).
The equivalence  follows from 
the fact that the identity of the canonical forms (\ref{canf}) implies that the corresponding 
Lagrangians and the Lagrangian actions for the both systems coincide
for the trajectories related by the canonical transformation $T$.
Hence, the solutions of these systems
are
 also related by this transformation since they are critical trajectories of the action.
 In other words, in the new variables, the Hamiltonian equations (\ref{canH})
 read similarly:
\be\la{canH2}
\dot \bA=D_\bPi \bH,\,\,\,\dot \bPi=-D_\bA \bH;\quad\dot q=D_P \bH,\,\,\,\dot P=-D_q \bH;
\quad\dot \vp=D_{M} \bH,\,\,\,\dot M=-D_\vp \bH.
\ee

Now the key point is that the Hamiltonian (\ref{HT}) 
does not depend on $q$ and $\vp$.  Hence, the equations for $q,P,\vp$, $M$ reduce to
\beqn\la{eqP}
\dot q&=&D_P\bH=\fr1{m} (P+\blangle \bPi, \na_*  \bA\brangle-\blangle \bA,\rho\brangle)
,\quad \dot P=0,
\\
\nonumber\\
 \dot \vp&=&D_M\bH=\fr1{I}(M+\blangle [Jy]\cdot \bA(y),\rho(y)\brangle)
 ,\qquad\quad\,\,\,\, \dot M=0.\la{eqM}
\eeqn
which also coincide with the formulas (\ref{Apv}), (\ref{moco}) and (\ref{pA})
and
correspond to the conservation of  $P$ and $M$. 
As the result, the system
 (\ref{canH2}) reduces to the following family of the 
Hamiltonian systems with parameters $P\in\R^2$ and  $M\in\R$:
\be\la{Hs2}
\dot \bA=D_{\bPi}\bH_{P,M},\quad \dot{\bPi}=-D_{\bA}\bH_{P,M},
\ee
where the ``reduced Hamiltonian''  is given by
\be\la{Hs3}
\bH_{P,M}(\bA,\bPi) =\bH(\bA,\bPi, P,M).
\ee
Expressing the variational derivatives, we write (\ref{Hs2}) as
\be\la{Hs4}
\left\{
\!\!\!\!\!\!
\ba{rcl}
\dot \bA&=&\bPi+
 (\dot q\cdot\na)\bA
\\
\quad \dot \bPi&=&\De \bA+
(\dot q \cdot\na)\bPi   +\cP[\dot q \rho(y)-\dot\vp Jy\rho(y)]
\ea\right|.
\ee
Here $\dot q$ and $\dot\vp$ are given by  (\ref{eqP}) and (\ref{eqM}).
 By the way,
these formulas allow us to reconstruct 
the entire  solution of the system (\ref{canH2}) knowing the solution of the reduced
system (\ref{Hs4}).
\setcounter{equation}{0}
\section{The solitons}\la{sA}
The solitons \eqref{solvom}
correspond to  stationary solutions
$\bS_{v,\om}=(\bA_{v,\om},\bPi_{v,\om})$
 of the reduced 
system (\ref{Hs4}) with the parameters $P_{v,\om}$
and $M_{v,\om}$ defined by  (\ref{eqP})--(\ref{eqM}):
\be\la{qpr2} 
 mv=P_{v,\om}+\blangle \bPi_{v,\om}, \na_*  \bA_{v,\om}\brangle
-\blangle  \bA_{v,\om},\rho\brangle,\qquad
I\om=M_{v,\om}+\blangle  [Jy]\cdot \bA_{v,\om}(y),\rho(y)\brangle.
\ee
    The stationary  equations corresponding to (\ref{Hs4}) read as
\be\la{Hs5}
0=\bPi_{v,\om}+
(v\cdot\na) \bA_{v,\om}, \qquad 0=\De \bA_{v,\om}+(v\cdot\na)\bPi_{v,\om}
+ \cP[v\rho(y)-\om Jy\rho(y)].
\ee
This means that 
the solitons $\bS_{v,\om}$ are critical points of the reduced
Hamiltonian $\bH_{P_{v,\om},M_{v,\om}}$.
We calculate $ \bA_{v,\om}$ from the first  equation of the system (\ref{2ml2}) with 
 $ A(x,t) = \bA_{v,\om}(x-v t)$:
\be\la{solit2}
\De  \bA_{v,\om}(y)-(v\cdot\na)^2 \bA_{v,\om}(y)=-{\cal P}[v\rho(y)]+\om Jy\rho(y).
\ee
The equation (\ref{solit2}) is easy to solve in the case $|v|<1$: in the Fourier transform
\be\la{solit3}
\hat  \bA_{v,\om}(k)
=\fr{(v-\fr{k(v\cdot k)}{k^2}+\om Ji\na)\hat \rho(k)}{k^2-(v\cdot k)^2}
\ee
by \eqref{Pi-e}. The conditions \eqref{rosym} and \eqref{rho02} imply that
\be\la{solH}
\bA_{v,\om}\in \dot H^1,\quad \bPi_{v,\om}=-(v\cdot\na) \bA_{v,\om}\in L^2.
\ee
Finally, equations  (\ref{eqP}) and (\ref{eqM}) with
$q(t)=vt$ and $\om(t)=\om$ hold according to  (\ref{qpr2}).
Thus, the solitons $\bS_{v,\om}\in \dot H^1\oplus L^2$ exist
for $(v,\om)\in\Si=\{(v,\om)\in\R^2\times\R:|v|<1 \}$.

The following lemma is proved in Appendix \ref{aJ}. 
It implies that the solitons $\bS_{v,\om}$ can be locally parametrised by their
total momenta $P_{v,\om}$ and angular momenta $M_{v,\om}$.
\bl\la{lJ}
The following Jacobian is nondegenerate:
\be\la{Jac}
J_{v,\om}=\det \fr{\pa(P_{v,\om},M_{v,\om})}{\pa(v,\om)}\ne 0,\qquad (v,\om)\in\Si.
\ee
\el
\setcounter{equation}{0}
\section{The Lyapunov function}
The  stationary  equations (\ref{Hs5}) mean that 
\be\la{nab0}
D_{ \bA} \bH_{P,M}( \bS_{v,\om})=0,\quad D_{\bPi} \bH_{P,M}( \bS_{v,\om})=0.
\ee
Hence, each  soliton $\bS_{v,\om}$ is a critical point of the reduced
Hamiltonian $\bH_{P,M}$ with the corresponding linear momentum $P=P_{v,\om}$ and 
angular momentum $M=M_{v,\om}$.
Moreover, it turns out that,  the
soliton is a strict global minimizer of $\bH_{P,M}$, which implies the orbital stability.
\begin{lemma} \la{llb}
The lower bound holds, 
\beqn\la{lo-bound} 
\de \bH_{P,M}&:=&\bH_{P,M}(\bA_{v,\om}+\de\bA,\bPi_{v,\om}+\de\bPi)-\bH_{P,M}(\bA_{v,\om},\bPi_{v,\om})
\nonumber\\
&\ge&\fr{1-|v|}{2}
(\Vert \de\bA\Vert_{\dot\cF^1}^2\!+\Vert \de\bPi\Vert_{\cF^0}^2),\qquad (\de\bA,\de\bPi)\!\in\! \cF.
\eeqn  
\end{lemma} 
\bpr Denote $\de\bA=\bba$, $\de\bPi=\bpi$. 
Then $\na\cdot \bba=0$, and  so we have 
\beqn \nonumber
&&\de \bH_{P,M}=
\fr12\int(|\bPi_{v,\om}+\bpi|^2+|\na (\bA_{v,\om}+\bba)|^2)\, dy
-\fr12\int(|\bPi_{v,\om}|^2+|\na \bA_{v,\om}|^2)\, dy\\
\nonumber
&&+\fr1{2m} (P+\blangle \bPi, \na_*  \bA\brangle-\blangle \bA,\rho\brangle)^2
-\fr1{2m} (P+\blangle \bPi_{v,\om}, \na_*  \bA_{v,\om}\brangle-\blangle \bA_{v,\om},\rho\brangle)^2\\
\la{dH}
&&+\fr1{2I}(M+\blangle[Jy]\cdot \bA,\rho\brangle)^2
-\fr1{2I}(M+\blangle [Jy]\cdot \bA_{v,\om},\rho\brangle)^2.
\eeqn
Evaluating, we obtain
\beqn\nonumber
&&\de \bH_{P,M}=
\fr12\int(|\bpi|^2+|\na \bba|^2)\, dy+
\int (\bPi_{v,\om}\cdot \bpi+\na\bA_{v,\om}\cdot\na\bba)\,dy\\
\nonumber
&&+\fr1{2m} \big(P+\!\blangle \bPi_{v,\om}, \na_*  \bA_{v,\om}\brangle
\!+\!\blangle \bpi, \na_*  \bA_{v,\om}\brangle
\!+\!\blangle \bPi_{v,\om}, \!\na_*  \bba\brangle
+\blangle \bpi, \na_*  \bba\brangle
\!-\!\blangle \bA_{v,\om},\rho\brangle\!-\!\blangle \bba,\rho\brangle\big)^2\\
\nonumber
&&
-\fr1{2m} \big(P+\blangle \bPi_{v,\om}, \na_*  \bA_{v,\om}\brangle-\blangle \bA_{v,\om},\rho\brangle\big)^2\\
\nonumber
&&+\fr1{2I}\big(M+\blangle[Jy]\cdot [\bA_{v,\om}+\bba],\rho\brangle\big)^2
-\fr1{2I}\big(M+\blangle [Jy]\cdot \bA_{v,\om},\rho\brangle\big)^2.
\eeqn
After a rearrangement, we get
\beqn \la{dH2}
\de \bH_{P,M}\!\!&\!\!=\!\!&\!\!
\fr12\int(|\bpi|^2+|\na \bba|^2)\, dy+
\int (\bPi_{v,\om}\cdot \bpi+\na\bA_{v,\om}\cdot\na\bba)\,dy
\nonumber\\
&&+\fr1{2m}[ (\p_{v,\om}+\de \p)^2
-\p_{v,\om}^2]
+\fr1{2I}[(M_{v,\om}+\de M)^2
-M_{v,\om}^2],
\eeqn
where
\be\la{pvb}
\left\{\ba{rcl}
\p_{v,\om}\!\!&\!\!:=\!\!&P\!\!+\blangle \bPi_{v,\om}, \na_*  \bA_{v,\om}\brangle-\blangle \bA_{v,\om}(y),\rho(y)\brangle=mv
\\
\de \p\!\!&\!\!:=\!\!&\!\!
\blangle \bpi, \na_*  \bA_{v,\om}\brangle
+\blangle \bPi_{v,\om}, \na_*  \bba\brangle
+\blangle \bpi, \na_*  \bba\brangle-\blangle \bba,\rho\brangle
\\
M_{v,\om}\!\!&\!\!:=\!\!&\!\!M+\blangle [Jy]\cdot \bA_{v,\om},\rho\brangle=I\om
\\
\de M\!\!&\!\!:=\!\!&\!\!\blangle [Jy]\cdot\bba,\rho\brangle
\ea\right|.
\ee
It remains to calculate the integral
$$
\cI:=\int (\bPi_{v,\om}\cdot \bpi+\na\bA_{v,\om}\cdot\na\bba)\,dy=
\blangle\bPi_{v,\om}, \bpi \brangle-\blangle  \De\bA_{v,\om},\bba\brangle.
$$
By (\ref{Hs5}), 
\be\la{rin2}
\cI=
-\blangle (v\cdot \na)\bA_{v,\om}, \bpi \brangle+\blangle  (v\cdot \na)\bPi_{v,\om},
\bba\brangle
+\blangle  v\rho-\om Jy\rho,\bba\brangle
\ee
since $\na\cdot \bba=0$.
Integrating by parts in (\ref{rin2}), we obtain
\be\la{rin3}
\cI=
-\blangle (v\cdot \na)\bA_{v,\om}, \bpi \brangle
-\blangle  \bPi_{v,\om},(v\cdot \na)\bba \brangle
+\blangle v\rho-\om Jy\rho,\bba\brangle
=-v\cdot\de \p-\om\de M+\blangle\bpi,(v\cdot \na)\bba\brangle.
\ee
Substituting into (\ref{dH2}), we get
\beqn\nonumber
\de \bH_{P,M}\!\!&\!\!=&\!\!
\fr12\int(|\bpi|^2+|\na \bba|^2)\, dy
+\blangle\bpi,(v\cdot\na)\bba\brangle\\
\la{dH3}
\!\!&\!\!\!\!&\!\!+\fr1{2m}[ (\p_{v,\om}+\de \p)^2
-\p_{v,\om}^2]-v\cdot\de \p
+\fr1{2I}[(M_{v,\om}+\de M)^2
-M_{v,\om}^2]-\om\de M.\qquad\qquad
\eeqn
The last line can be written as
\beqn\nonumber
&&\fr m{2}\Big( (v+\de \p/m)^2-v^2-2v\cdot\de \p/m\Big)
+\fr I{2}\Big((\om+\de M/I)^2-\om^2-2\om\de M/I\Big)\\
&&=\fr m{2}(\de \p/m)^2+\fr I{2}(\de M/I)^2\ge 0.
\la{LL}
\eeqn
We have $|\blangle\bpi,(v\cdot\na)\bba\brangle|\le|v||\bpi||\na\bba|$.
Hence, \eqref{dH3} and (\ref{LL}) imply that
\beqn\nonumber
\de \bH_{P,M}&\ge&\fr12\int(|\bpi|^2+|\na \bba|^2)\, dy
+\blangle\bpi,(v\cdot\na)\bba\brangle\ge \frac 12\int(|\bpi|^2-2|v||\bpi||\na\bba|+|\na\bba|^2)dy\\
\nonumber
&=&\frac{1-|v|}2\int(|\bpi|+|\na\bba|)^2dy+\frac{|v|}2\int(|\bpi-\na\bba||^2dy\ge\frac{1-|v|}2\int(|\bpi|+|\na\bba|)^2dy. 
\eeqn

 \epr 
\setcounter{equation}{0}
\section{The orbital stability of  solitons} 
Introduce the distance between two solutions $\bF_1(y,t)$ and $\bF_2(y,t)$  as
\be\la{dist}
 d( \bF_1(t), \bF_2(t)):=\Vert \bF_1(t)-\bF_2(t)\Vert_{\cF}
+|\dot q_1(t)-\dot q_2(t)|+|\dot \vp_1(t)-\dot \vp_2(t)|, \qquad t\in\R,
\ee
where $\dot q_j(t)$ and $\dot \vp_j(t)$ are defined by (\ref{eqP})--(\ref{eqM}).
Recall that the reduced soliton (\ref{solvom})  does not depend on time:
\be\la{refia}
 \bF_{v,\om}(y,t)=\bS_{v,\om}(y):=(\bA_{v,\om},\bPi_{v,\om}).
\ee
\bd
The soliton $\bS_{v,\om}$  is orbitally stable if  for any $\ve>0$ there is 
a $\de>0$ such that 
for any solution $\bF$ of the system (\ref{Hs4}),
the inequality 
\be\la{tYt}
d(\bF(0),\bS_{v,\om})<\de
\ee
implies
\be\la{tYt2}
 d( \bF(t), \bS_{v,\om})<\ve,\qquad t\in\R.
\ee
\ed
\bt\la{t1}
Let $|v|<1$ and let $\rho(x)$ satisfy the conditions {\rm (\ref{rosym})} and {\rm (\ref{rho02})}. 
Then  the soliton {\rm (\ref{solvom})} is orbitally stable. 
\et
\bpr
For the proof we develop  the general scheme of \ci{BG1993, IKM2004}.
Namely, by Lemma \ref{llb},
the soliton $\bS_{v,\om}$ is a nondegenerate 
global minimizer of the reduced Hamiltonian
$\bH_{P,M}$.
The inequality (\ref{tYt}) with sufficiently small $\de>0$
implies that the linear and angular momenta $P^*$ and $M^*$ of the solution
$\bF(t)$ are close to $P$ and $M$, respectively:
\be\la{mclo}
 |P^*-P|+|M^*-M| \to 0,\qquad \de\to 0.
\ee
By Lemma \ref{lJ}, for sufficiently small $\de>0$, there exists a unique
soliton $\bS_{v_*,\om_*}$
with the total momentum $P^*$ and angular momentum $M^*$,
which
 is close to $\bS_{v,\om}$:
\be\la{crp}
\Vert \bS_{v_*,\om_*}-\bS_{v,\om}\Vert_{\cF}
\to 0,
\qquad\de\to 0.
\ee
Combining with  (\ref{tYt}), we get
\be\la{crp2}
\Vert \bF(0)-\bS_{v_*,\om_*}\Vert_{\cF}\to 0,\qquad\de\to 0.
\ee
Therefore, 
\be\la{crp3}
|\bH_{P_*,M_*}(\bF(0))-\bH_{P_*,M_*}(\bS_{v_*,\om_*})|\le \de_1,
\qquad \de_1\to 0\,\,\,{\rm as}\,\,\,\de\to 0.
\ee
Now
the 
conservation of the reduced Hamiltonian $\bH_{P_*,M_*}$ implies that 
\be\la{crp4}
|\bH_{P_*,M_*}(\bF(t))-\bH_{P_*,M_*}(\bS_{v_*,\om_*})|\le \de_1,\quad t\in\R.
\ee
Therefore, the bound (\ref{lo-bound})
with $v,\om$ replaced by $v_*,\om_*$,
allows us to conclude that
\be\la{crp5}
\sup_{t\in\R}
\Vert \bF(t)-\bS_{v_*,\om_*}\Vert_{\cF}\to 0\,\,\,{\rm as}\,\,\,\de\to 0.
\ee
Finally, combined with  (\ref{crp}), this inequality  gives 
\be\la{crp6}
\sup_{t\in\R}
\Vert \bF(t)-\bS_{v,\om}\Vert_{\cF}\to 0\,\,\,{\rm as}\,\,\,\de\to 0.
\ee

It remains to show that
the terms with $\dot q$ and $\dot \vp$ in (\ref{dist}) are small uniformly in time:
\be\la{ter}
|\dot q(t)-v|+|\dot \vp(t)-\om|\le |\dot q(t)-v_*|+|\dot \vp(t)-\om_* |+|v_*-v|
+|\om_*-\om|.
\ee
Indeed, the last two terms are  small by   (\ref{qpr2}) together with
(\ref{mclo}) and (\ref{crp}).  
Similarly,
the smallness of the first two terms follows from  (\ref{eqP})--(\ref{eqM})
together with (\ref{mclo})  and (\ref{crp5}).
\epr
\setcounter{equation}{0}
\section{The case of a relativistic particle}
In this section we comment on the required modifications of our arguments in the case
of a relativistic particle with kinetic  momentum
$$
\p=m\dot q/\sqrt{1-\dot q^2}.
$$
Now  the third equation of the system (\ref{mls2}) should be replaced by
$$
\dot \p(t)=\blangle E(x,t)+[J\dot q(t) +\om(t) (x-q(t)) ]B(x,t) ,\rho(x\!-\!q(t))\brangle.
$$
Respectively, the second equation of the system (\ref{2ml2}) now reads
$$
\dot \p(t)=\blangle-\dot A(x,t)+[J\dot q(t) +\om(t) (x-q(t))),\rho(x\!-\!q(t)]\na\cdot JA(x,t) \brangle.
$$
The Lagrangian (\ref{LagRA}) is replaced by
$$
 L(A,\dot A,q,\dot q,\vp,\dot\vp)=
\fr12\int[\dot A^2-(\na A)^2]dx- m\sqrt{1-\dot q^2}+\fr{I\dot \vp^2}2
+\blangle [\dot q-\dot\vp J(x-q)]\cdot A,\rho(x-q)\brangle. 
$$
Accordingly, the momentum in (\ref{pA}) changes to
\be\la{pAr}
p=\p+\blangle \bA,\rho\brangle.
\ee
The equations (\ref{eq2}) and (\ref{mdo})
become
\beqn\nonumber
&&\dot \p_k\!+\!\blangle\dot A_k(q\!+\!y,t),\rho(y)\brangle+\blangle(\dot q(t)\!\cdot\!\na) A_k(q\!+\!y,t),
\rho(y)\brangle
\!=\!\blangle[\dot q\!-\!\dot\vp Jy]\cdot\na_k A(q\!+\!y,t),\rho(y)\brangle,\\
\nonumber 
&&\dot\p_k(t)=\blangle-\dot A_k(q+y,t)+
[(J\dot q)_k(t) +\dot\vp(t) y_k]
\na\cdot JA(q+y,t),\rho(y) \brangle,\quad k=1,2.
\eeqn
Now the linear momentum  (\ref{moco}) keeps its form with $p$ given by (\ref{pAr}),
and hence the
formula for $\dot q$ in (\ref{Apv}) is replaced by
\be\la{dqu}
\dot q=\fr{\p}{\sqrt{m^2+\p^2}}=\fr{ p-\blangle \bA,\rho\brangle}
{\sqrt{m^2+(p-\blangle \bA,\rho\brangle)^2}}
=\fr{P+\blangle \bPi,\na_*\bA\brangle-\blangle \bA,\rho\brangle}
{\sqrt{m^2+(P+\blangle \bPi,\na_*\bA\brangle-\blangle \bA,\rho\brangle)^2}}.
\ee
The Hamiltonian (\ref{Hp}) becomes
\be\la{Hpr}
H=\fr12\int[\Pi_A^2\!+\!|\na A|^2]dx\!+\!\fr {m}{\sqrt{1\!-\!\dot q^2}}\!+\!\fr{I\dot \vp^2}2.
\qquad\qquad
\ee
Here  $m/\sqrt{1-\dot q^2}$  is expressed via  (\ref{dqu}):
\be\la{dqu2}
\fr {m}{\sqrt{1-\dot q^2}}=\sqrt{m^2+\p^2}:=\sqrt{m^2+(P+\blangle \bPi,\na_*\bA\brangle-\blangle \bA,\rho\brangle)^2},
\ee
while $I\dot \vp^2/2$ is given by  (\ref{Apv})
 as for the nonrelativistic particle.

The system (\ref{Hs4}) remains the same, but now $\dot q$ is defined by (\ref{dqu}),
while $\dot\vp$ remains (\ref{eqM}). As a result,  equations (\ref{Hs5}) for the soliton fields  are the same. The difference
 is that in the relativistic case the inequality $|v|<1$ holds automatically.
\smallskip

Finally, Lemma \ref{llb} remains valid in the case of a relativistic particle as well. The changes in the proof are as follows.
The second line of (\ref{dH}) is replaced by
$$
\sqrt{m^2+\p^2}-\sqrt{m^2+\p_v^2}=\sqrt{m^2+(\p_v+\de \p)^2}-\sqrt{m^2+\p_v^2}.
$$
Here $ \p_v:=mv/\sqrt{1-v^2}$ and
$$
\de\p:=\p-\p_v=
(P+\blangle \bPi,\na_*\bA\brangle-\blangle \bA,\rho\brangle)
-(P+\blangle \bPi_{v,\om},\na_*\bA_{v,\om}\brangle-\blangle \bA_{v,\om},\rho\brangle),
$$
which coincides with the second line of (\ref{pvb}). Hence,
the formula (\ref{rin3}) remains valid, and so (\ref{dH3}) becomes 
\beqn \la{dH3r}
\de \bH&=&
\fr12\int(|\bpi|^2+|\na \bba|^2)\, d^3x
+\blangle\bpi,d_v\bba\brangle
\nonumber\\
&&
+\sqrt{m^2+(\p_v+\de \p)^2}-\sqrt{m^2+\p_v^2}
-v\cdot\de \p
+\fr I{2}(\de M/I)^2.
\eeqn
Here
\be\la{lalr}
\sqrt{m^2+(\p_v+\de \p)^2}-\sqrt{m^2+\p_v^2}
-v\cdot\de \p\ge 0
\ee
since the function $E(\p):=\sqrt{m^2+\p^2}$ is convex 
(its graph is the hyperbola $E^2-\p^2=m^2$)
and $\na E(\p_v)=v$. 
Hence, (\re{lo-bound}) follows.
\smallskip

Finally, Theorem \ref{t1} and its proof remain unchanged.

\appendix
\section{On alternative paramerisation of solitons}\la{aJ}
Here we prove Lemma \ref{lJ}.
 The Jacobian (\ref{Jac}) reads
\be\la{Jac2}
J_{v,\om}=\det\left(\!\!\!
\ba{ccc}
\fr{\pa P_1}{\pa v_1}&\fr{\pa P_2}{\pa v_1}&\fr{\pa M}{\pa v_1}
\\
\fr{\pa P_1}{\pa v_2}&\fr{\pa P_2}{\pa v_2}&\fr{\pa M}{\pa v_2}
\\
\fr{\pa P_1}{\pa \om}&\fr{\pa P_2}{\pa \om}&\fr{\pa M}{\pa \om}
\ea
\!\!\!\right); \quad M=M_{v,\om}; \quad P_j=(P_{v,\om})_j,\quad j=1,2.
\ee
By (\ref{qpr2}),
\be\la{Mv}
\fr{\pa M}{\pa \om}=I+\int\frac{|\na\hat\rho|^2dk}{k^2-(v\cdot k)^2}>0 ,\quad
\fr{\pa M}{\pa v_l}=2\om\int\frac{(v\cdot k)k_l|\na\hat\rho|^2}{(k^2-(v\cdot k)^2)^2} dk,\quad l=1,2.
\ee
Further,
we split $\bA=\bA_{v,\om}$  into even  ``+" and odd ``-" components. 
In the Fourier transform,
$\hat\bA=\hat\bA^++\hat\bA^-$, where, by (\ref{solit3}),
$$
\hat\bA^+(k)
=\ds\fr{(v-\fr{k(v\cdot k)}{k^2})\hat \rho(k)}{k^2-(v\cdot k)^2},
\qquad
\hat\bA^-(k)
=\ds\fr{i\om J\na\hat \rho(k)}{k^2-(v\cdot k)^2}.
$$
Respectively, using (\ref{qpr2}), we split  $P$  as  $P=P^++P^-$, where
$$
P^-_j=\langle (v\cdot \na) \bA^-,\na_j\bA^-\rangle=\langle (v\cdot k) \hat\bA^-,k_j\hat\bA^-\rangle
=\om^2\int\frac{(v\cdot k) k_j|\na\hat\rho|^2dk}{(k^2-(v\cdot k)^2)^2}, 
$$
\beqn\nonumber
 P^+_j\!\!&=&\!\!mv_j+\blangle  \bA^+_j,\rho\brangle+\langle (v\cdot\na)\bA^+, \na_j  \bA^+\rangle
 =mv_j+\blangle \hat\bA^+_j,\hat\rho\brangle+\langle (v\cdot k)\hat\bA^+, k_j  \hat\bA^+\rangle
 \\
 \nonumber
 \!\!&=& \!\!mv_j+v_j\!\int\frac{|\hat\rho|^2dk}{k^2\!-\!(v\cdot k)^2}-\!\int\frac{(v\cdot k)k_j|\hat\rho|^2dk}{k^2(k^2\!-\!(v\cdot k)^2)} 
+ \int\big(v^2\!-\!\fr{(v\cdot k)^2}{k^2}\big)\frac{(v\cdot k) k_j|\hat\rho|^2dk}{(k^2\!-\!(v\cdot k)^2)^2}. 
\eeqn
Differentiation with respect to $v_l$ and $\om$ gives
\beqn\la{P-}
\fr{\pa P^-_j}{\pa v_l}&=&\om^2\!\int\frac{k_jk_l|\na\hat\rho|^2dk}{(k^2-(v\cdot k)^2)^2}
+\om^2\!\int\frac{4(v\cdot k)^2 k_jk_l|\na\hat\rho|^2dk}{(k^2-(v\cdot k)^2)^3}\\   
\nonumber
\fr{\pa P^+_j}{\pa v_l}\!\!&=&\!\!\delta_{jl}\Big(m+\int\frac{|\hat\rho|^2dk}{k^2-(v\cdot k)^2}\Big)+v_j\int\frac{2(v\cdot k)k_l|\hat\rho|^2dk}{(k^2-(v\cdot k)^2)^2}
-\int\frac{k_jk_l|\hat\rho|^2dk}{k^2(k^2-(v\cdot k)^2)}\\
\nonumber
\!\!&-&\!\!\int\frac{2k_jk_l(v\cdot k)^2|\hat\rho|^2}{k^2(k^2-(v\cdot k)^2)^2} dk
+\int\big(v^2-\fr{(v\cdot k)^2}{k^2}\big)\frac{k_jk_l|\hat\rho|^2dk}{(k^2-(v\cdot k)^2)^2}\\
\la{P+}
\!\!&+&\!\!\int\big(v_l-\fr{(v\cdot k)k_l}{k^2}\big)\frac{2(v\cdot k) k_j|\hat\rho|^2dk}{(k^2-(v\cdot k)^2)^2}
+\int\big(v^2-\fr{(v\cdot k)^2}{k^2}\big)\frac{4(v\cdot k)^2 k_jk_l|\hat\rho|^2dk}{(k^2-(v\cdot k)^2)^3},\\
\la{P-om}
\fr{\pa P_j}{\pa \om}&=&\fr{\pa P^-_j}{\pa \om}=2\om\!\int\frac{(v\cdot k) k_j|\na\hat\rho|^2dk}{(k^2-(v\cdot k)^2)^2}.
\eeqn

We can assume that $v=(|v|,0)$.  In this case $\ds\fr{\pa P_2}{\pa \om}=0$ and 
$\ds\fr{\pa P_j^+}{\pa v_l}=\ds\fr{\pa P_j^-}{\pa v_l}=0$ for $j\ne l$. Hence, \eqref{Jac2}  becomes
\be\la{Jac3}
J_{v,\om}=\fr{\pa P_2}{\pa v_2}\Big(\fr{\pa P_1}{\pa v_1}\fr{\pa M}{\pa \om}-\fr{\pa M}{\pa v_1}\fr{\pa P_1}{\pa \om}\Big).
\ee
Denoting $D=k^2-v^2k_1^2$,  \eqref{P-} and \eqref{P+} gives
\be\la{Pj-}
\fr{\pa P^-_j}{\pa v_j}=\om^2\int\frac{k_j^2|\na\hat\rho|^2dk}{D^2}
+\om^2\int\frac{4v^2k_1^2 k_j^2|\na\hat\rho|^2dk}{D^3}\ge 0, \quad 
\fr{\pa P_1}{\pa \om}=2\om|v|\!\int\frac{k_1^2|\na\hat\rho|^2dk}{D^2},
\ee
\beqn\nonumber
\fr{\pa P^+_1}{\pa v_1}
\!\!&=&\!\!m+\int\frac{|\hat\rho|^2dk}{D}+\int\frac{2v^2k_1^2|\hat\rho|^2}{D^2} dk
-\int\frac{k_1^2|\hat\rho|^2}{k^2D} dk-\int\frac{2v^2k_1^4|\hat\rho|^2}{k^2D^2} dk\\
\nonumber
\!\!&+&\!\!\int\frac{v^2k_1^2k_2^2|\hat\rho|^2}{k^2D^2}+\int\frac{2v^2k_1^2k_2^2|\hat\rho|^2}{k^2D^2}
+\int\frac{4v^4k_1^4k_2^2|\hat\rho|^2}{k^2D^3}\\
\la{P1+}
\!\!&=&\!\!m+\int\frac{k_2^2|\hat\rho|^2dk}{k^2D}
+\int\frac{5v^2k_1^2k_2^2|\hat\rho|^2}{k^2D^2}+\int\frac{4v^4k_1^4k_2^2|\hat\rho|^2}{k^2D^3}>0.
\eeqn
\beqn\nonumber
\fr{\pa P^+_2}{\pa v_2}
\!\!&=&\!\!m+\int\frac{|\hat\rho|^2dk}{D}
-\int\frac{k_2^2|\hat\rho|^2dk}{k^2D} -\int\frac{2v^2k_2^2k_1^2|\hat\rho|^2dk}{k^2D^2} \\
\nonumber
\!\!&+&\!\!\int\frac{v^2k_2^4|\hat\rho|^2dk}{k^2D^2}
-\int\frac{2v^2k_1^2k_2^2|\hat\rho|^2dk}{k^2D^2}+\int\frac{4v^4k_1^2k_2^4|\hat\rho|^2dk}{k^2D^3}\\
\la{P2+}
\!\!&=&\!\!m+\int\frac{v^2k_2^4|\hat\rho|^2dk}{k^2D^2}
+\int\frac{k_1^2|\hat\rho|^2}{k^2D}\Big(1-\frac{4v^2k_2^2}{D}+\frac{4v^4k_2^4)}{D^2}\Big) dk>0
\eeqn
since $1-\ds\frac{4v^2k_2^2}{D}+\ds\frac{4v^4k_2^4}{D^2}=\Big(1-\ds\frac{2v^2k_2^2}{D}\Big)^2$.
From \eqref{Pj-} and \eqref{P2+} it follows that  
\be\la{P2}
\fr{\pa P_2}{\pa v_2}=\fr{\pa P_2^+}{\pa v_2}+\fr{\pa P_2^-}{\pa v_2}>0.
\ee
Moreover,  \eqref{Mv} and \eqref{Pj-}--\eqref{P1+}   imply that
\beqn\nonumber
\fr{\pa P_1}{\pa v_1}\fr{\pa M}{\pa\om}-\fr{\pa M}{\pa v_1}\fr{\pa P_1}{\pa \om}&=&\fr{\pa P_1^+}{\pa v_1}\fr{\pa M}{\pa\om}
+\fr{\pa P_1^-}{\pa v_1}\fr{\pa M}{\pa\om}-\fr{\pa M}{\pa v_1}\fr{\pa P_1}{\pa \om}
=\fr{\pa P_1^+}{\pa v_1}\fr{\pa M}{\pa\om}\\
\nonumber
\!\!&+&\!\!
\om^2 I\Big(\int\frac{k_1^2|\na\hat\rho|^2dk}{D^2}+\!\int\frac{4v^2k_1^4|\na\hat\rho|^2dk}{D^3}\Big)
+\om^2\int\frac{|\na\hat\rho|^2dk}{D}\cdot\!\int\frac{k_1^2|\na\hat\rho|^2dk}{D^2}
\\
\la{PPMM}
\!\!&+&\!\!4\om^2v^2\Big(\int\frac{|\na\hat\rho|^2dk}{D}\cdot\int\frac{k_1^4|\na\hat\rho|^2dk}{D^3}
-\Big(\int\frac{k_1^2|\na\hat\rho|^2dk}{D^2}\Big)^2\Big)>0
\eeqn
since $\ds\fr{\pa P_1^+}{\pa v_1}\fr{\pa M}{\pa\om}>0$ by \eqref{Mv} and \eqref{P1+}, and
$
\ds\Big(\int\frac{k_1^2|\na\hat\rho|^2dk}{D^2}\Big)^2\le \int\frac{|\na\hat\rho|^2dk}{D}\cdot\int\frac{k_1^4|\na\hat\rho|^2dk}{D^3}
$
by the Cauchy-Schwarz inequality. As the result, $J_{v,\om}>0$ by \eqref{Jac3}, \eqref{P2} and  \eqref{PPMM}. 


\begin{thebibliography}{99}


\bibitem{A1989}
V. I. Arnold, Mathematical Methods of Classical Mechanics, Springer, New York, 1989.

\bibitem{BG1993}
D. Bambusi, L. Galgani,
Some rigorous results on the Pauli--Fierz model of classical electrodynamics, 
{\em Ann. de l'I.H.P., section A} {\bf 58} (1993), no. 2, 155--171. 

\bibitem{HK2000}
G. Hoermann, M. Kunzinger,
Regularized derivatives in a 2-dimensional model of self-interacting fields with singular data,
{\em Zeitschr. Anal. Anw.} {\bf 19} (2000) 1, 147--158.

\bibitem{H1974}
G. 't Hooft,
A two-dimensional model for mesons, Nuclear Physics B75 (1974), 461--470.

\bibitem{IKM2004}
V. Imaykin, A. Komech, N. Mauser, 
Soliton-type asymptotics for the coupled Maxwell--Lorentz equations, 
{\em Ann. Inst. Poincar\'e, Phys. Theor.} {\bf  5} (2004) 1117--1135.

\bibitem{IKS2004}
V. Imaykin, A. Komech, H. Spohn, 
Rotating charge coupled to the Maxwell field: scattering theory and adiabatic limit, 
{\em Monatsh. Math.} {\bf 142} (2004), no. 1--2,  143--156.
 
 \bibitem{IKS2011}
V. Imaykin, A. Komech, H. Spohn, 
Scattering asymptotics for a charged particle coupled to the Maxwell field, 
{\em J. Math. Phys.} {\bf 52} (2011), no. 4,  042701.

\bibitem{KK2020}
A. Komech, E. Kopylova,
Attractors of  nonlinear Hamiltonian partial differential equations, 
 {\em Russ. Math. Surv.} {\bf 75} (2020), no. 1,  1--87.

\bibitem{KK2022}
A. Komech, E. Kopylova,
Attractors of  Hamiltonian Nonlinear Partial Differential Equations, 
Cambridge University Press, Cambridge, 2022.

\bibitem{KK2023}
A. Komech, E. Kopylova,
On the stability of solitons for Maxwell--Lorentz equations with rotating particle, 
{\em Milan Journal of Mathematics} {\bf 91} (2023),  155--173.



\bibitem{KK2023_LH}
A. Komech, E. Kopylova,
On the Hamilton--Poisson structure and solitons for the Max\-well--Lo\-rentz 
equations with spinning particle,  {\em J. Math. Anal. Appl.} {\bf 522} (2023),
no. 2,  126976.


\bibitem{KK2024}
A. Komech, E. Kopylova,
On orbital stability of solitons  for 3D Maxwell--Lorentz 
equations with spinning particle,
 arXiv:2306.00508 [math-ph].



\bibitem{KS2000}
A.I. Komech, H. Spohn, 
Long-time asymptotics for the coupled Maxwell--Lorentz equations, 
{\em Comm. Partial Differential Equations} {\bf 25} (2000), 559--584.

\bibitem{K2010}
M. Kunze, 
On the absence of radiationless motion for a rotating classical charge
{\em Advances in Mathematics} {\bf 223}, no. 5,  1632--1665.

\bibitem{LL1975}
L.D. Landau, E.M. Lifshitz, The classical theory of fields, Pergamon, 1975.

\bibitem{MR2002}
J.E. Marsden, T. Ratiu, 
Introduction to Mechanics and Symmetry, Springer, New York, 2002.

\bibitem{S1962}
J. Schwinger, 
Gauge invariance of mass II, Phys. Rev. 128 (1962), no. 5, 2425--2429.

\bibitem{S2004}
 H. Spohn, Dynamics of Charged Particles and Their Radiation Field, 
Cambridge University Press, Cambridge, 2004.


\end{thebibliography}
\end{document}